\documentclass[aps,prd,twocolumn,showpacs,nofootinbib]{revtex4}
\usepackage{latexsym}
\usepackage{amsmath,amsfonts}
\usepackage{amsbsy}
\usepackage{mathrsfs}
\usepackage{color}

\usepackage{psfrag}

\usepackage{enumerate}

\usepackage{amsmath,amssymb,calc,amsfonts}
\usepackage{latexsym}

\usepackage{graphicx,calc,epsfig}
\def\ut#1{\rlap{\lower1ex\hbox{$\sim$}}#1{}}
\newcommand{\N}{\mathbb{N}}

\newcommand{\be}{\nopagebreak[3]\begin{equation}}
\newcommand{\ee}{\end{equation}}
\newcommand{\ba}{\nopagebreak[3]\begin{eqnarray}}
\newcommand{\ea}{\end{eqnarray}}
\DeclareFontFamily{U}{rsfs}{}         % Formal Script            %
\DeclareFontShape{U}{rsfs}{m}{n}{<5> rsfs5 <6><7> rsfs7          %
  <8><9><10><10.95><12><14.4><17.28><20.74><24.88> rsfs10}{}     %
\DeclareMathAlphabet{\mathfs}{U}{rsfs}{m}{n}                     %
\newcommand{\mfs}[1]{\mathfs {#1}}                               %
\newcommand{\sA}{{\mfs A}}

\newcommand{\sW}{{\mfs W}}
\newcommand{\sO}{{\mfs O}}

\def\pb#1{\rlap{\lower1.5ex\hbox{$\longleftarrow$}}{#1}}
\def\dpb#1{\rlap{\lower1.5ex\hbox{$\Longleftarrow$}}{#1}}
\def\spb#1{\rlap{\lower1.5ex\hbox{$\leftarrow$}}{#1}}
\def\sdpb#1{\rlap{\lower1.5ex\hbox{$\Leftarrow$}}{#1}}

\definecolor{blue}{rgb}{0,0,1}
\definecolor{green}{rgb}{0,1,0}
\definecolor{red}{rgb}{1,0,0}
\definecolor{vio}{rgb}{1,0,1}
\definecolor{ama}{rgb}{1,1,0}

\begin{document}

% \affiliation{$^2$Centre de Physique Th\'eorique\footnote{Unit\'e
% Mixte de Recherche (UMR 6207) du CNRS et des Universit\'es
% Aix-Marseille I, Aix-Marseille II, et du Sud Toulon-Var; laboratoire
% afili\'e \`a la FRUMAM (FR 2291)}, Campus de Luminy, 13288
% Marseille, France.}
%
%
% \affiliation{$^1$ \blue{FaMAF, Universidad Nacional de C\'ordoba,
%  Instituto de F\'isica Enrique Gaviola (IFEG), CONICET, \\
%  Ciudad Universitaria, (5000) C\'ordoba, Argentina.}
% }

%\maketitle

%\begin{abstract}
%\end{abstract}

%%\pacs{}
\title{Reply to the comment on ``Black hole entropy and isolated horizons thermodynamics"}

\date{\today}

\author{Amit Ghosh$^1$}
\author{Alejandro Perez$^2$}

\affiliation{$^1$Saha Institute of Nuclear Physics, 1/AF Bidhan Nagar,
700064 Kolkata, India.\\
$^2$Centre de Physique Th\'eorique, Campus de Luminy, 13288
Marseille, France.}

%%
%%\affiliation{$^1$ . }

\vskip-3cm
\begin{abstract}
The comment 	arXiv:1204.2729v1 is completely wrong.
The author makes serious mistakes
in calculations and judgement. The errors are made at the
level of basic undergraduate statistical mechanics.
%His comment is simply empty.
\end{abstract}

%\pacs{}

\maketitle

%\noindent Dear Editor:

%We apologize to the reader of this prestigious journal for having to use such valuable space to address a completely misplaced criticism which is wrong at such a basic level that, in our humble opinion, should have never been accepted for publication in any journal.

%As we carefully show in what follows there is no error in our letter but only in A. Majhi's calculation
%and understanding.
This is a reply to a comment that A. Majhi that appeared recently \cite{idiot}. The present reply addresses the points as raised in version arXiv:1204.2729v1 of the paper (the only version available at the moment of writing this manuscript).
% This reply is adapted to address the points raised in an extended version 
% published on the arxiv \cite{idiot}.
The comment contains two issues appearing in the first page.
Our reply to these points are to be found in the following section.
The rest of the comment \cite{idiot} contains further issues here addressed in  appendix \ref{more}.

\section{reply to the comments}\label{uno}

{\bf On ``Issue 1":} We study a system of $N$ distinguishable
particles having single-particle energy levels $E_j=\gamma \ell_p^2/\ell
\sqrt{j(j+1)}$ where $j\in \N/2$. We perform the analysis in the microcanonical,
canonical and (briefly at the end) the grand canonical ensembles.
We show that all the three treatments are equivalent in the appropriate regime.
There is only one physical input (well justified in our
letter): the system represents a black hole only if it is in thermal
equilibrium at the Unruh temperature
$T_U=\ell_p^2/(2\pi \ell)$. 

In the microcanonical ensemble treatment of \cite{us} the is a Lagrange multiplier called
$\lambda$ which is basically the inverse temperature (in text books $\lambda\propto\beta$; we used $\lambda$ just to keep the notation similar to previous literature). 
So choosing a value for
$\lambda$ is completely equivalent to choosing a temperature
$(\partial S/\partial E|_N)^{-1}$, here $T_U$ (in a similar way a temperature
can be chosen for an ideal gas and the results are in complete analogy with
ours). There is no problem at this level this is just standard statistical mechanics.

From here A. Majhi goes on and attempts to construct a new argument which is wrong at the calculus level from which he develops an equally wrong conceptual conclusion. Let us see this in detail.

First the calculational error: A. Majhi correctly notices that the second multiplier $\sigma$, which is simply related to the chemical potential, appears in the equation ${\sA}/{N}=-{d\sigma}/{d\lambda}$. Then he makes a basic mistake concluding that this implies
$(\lambda-\lambda_0){\sA}/{N}=\sigma(\lambda_0)-\sigma(\lambda).$ This would be true only if ${d\sigma}/{d\lambda}={\rm constant}$ which
is clearly not the case (from (14) in \cite{us} $\sigma(\lambda)=\log[\sum
(2j+1)\exp(-\lambda\sqrt{j(j+1)})]$).
%(since $\lambda$ is a function of $\sA$ and $N$).
%(see eq. 14 in our letter).
In fact, integrating does not give new information. However, setting $T=T_U$, or equivalently $\lambda=2\pi\gamma$, 
does:
\be\label{one}
\frac{\sA}{N}=-\frac{d\sigma}{d\lambda}\ \ \ \Longrightarrow\ \ \ \frac{\sA}{N}=-\frac{1}{2\pi}\frac{d\sigma}{d\gamma}
\ee
which is precisely  eq. (19) in \cite{us}. Thus, had A. Majhi
not made the mistake he would have got the microcanonical derivation of (19)
(in \cite{us} we used the canonical ensemble). 
%In fact, if one blindly follows the calculations of A. Majhi, one can prove a little theorem that for all systems entropy is independent of temperature! It is wrong at this level. 

The second, more important, error is conceptual: eq. (\ref{one}) does not
contradict the independence of $\sA$ and $N$ as thermodynamic variables. It is
a relation between $\sA$ and $N$ at thermal equilibrium. 
The analogous relation for an ideal Boltzmann gas is $E=\frac{3}{2}NT$, which at 
any fixed temperature (like
$T=T_U$), relates $E$ (the analogue of
$\sA$) to $N$. It is like an equation of state---by no means it implies that $E$
and $N$ cannot be treated as independent thermodynamic variables for an ideal
gas.

% Using the wrong version of it A. Majhi concludes that one can express
% $\lambda=f(\sA/N)$. Even when the form of his $f$ is incorrect one could use
% eq. (\ref{one}) above to establish a similar relationship. This simply says that
% $\lambda$ (or equivalently the temperature as the two are related (see text below eq. (14) in \cite{us}))
% is a function of $\sA/N$ just as $T=U/(\frac{2}{3}kN)$ for an ideal gas.

{\bf On ``Issue 2'':}
Here one is dealing with the same system in the {\em grand canonical ensemble}.
The grand canonical parition function is $\mfs Z=(1-zf(T))^{-1}$. The main
worry of A. Majhi is that $\mfs Z$ is divergent because $zf(T)=1$ and also
how did we get $zf(T)=1+o(1/N)$? Again, his mistake is elementary and
can easily be sorted out as we show now.
The average number of particles is $\langle N\rangle=z(\partial/\partial
z)\ln\mfs Z= zf(T)(1-zf(T))^{-1}$. Solving this
%\be
$zf(T)={\langle N\rangle}/({\langle N\rangle+1})$,
%\ee
and $\mfs Z=1+\langle N\rangle$. This eliminates the erroneous divergence that
A. Majhi finds. In fact it should be clear that this
form of $\mfs Z$ does not depend on any particular single-particle spectrum and
%Hence the form is fairly general and no special pathology occurs for our
%spectrum.
is so general that such calculations appear in many
basic text books (e.g. see \cite{pa}).
% (we encourage A. Majhi
%to read it).

Only in the canonical and microcanonical
ensembles $zf(T)=1$ at $T=T_U$. In fact if any conclusion is sought in the grand
canonical
ensemble, one should compute things in the {\em grand canonical ensemble}
(to copy expressions obtained in the canonical
ensemble and use them blindly in the grand canonical ensemble is not correct).

{\bf Summary:} We have shown that every single issue raised by A. Majhi's is
based on either a wrong calculation or confusion of the most basic
concepts in
statistical mechanics and thermodynamics.

%This concludes our reply.

To end in a more positive tone we briefly review the core of the result presented in \cite{us}.
%(it is
%unfortunate that our solid results received such vapid and vacuous attacks of A.
%Majhi which unavoidably induces unfounded suspicions on the minds of careless
%readers):
There are two expressions for the entropy
% at thermal equilibrium,
(analogous to the Sackur-Tetrode equation for an ideal gas), namely
\be\label{esta}
S=\frac{A}{4\ell^2_p}+\sigma(\gamma) N,  \ \  \ \mbox{or}\ \ \  S=\frac{A}{4\ell^2_p}(1-\frac{\sigma}{\gamma \sigma^{\prime}}).\ee The semiclassical consistency of
this entropy (which does depend on $\gamma$ in a non trivial way) comes from the
fact that  (for stationary black holes)
\be \label{pri}\delta M=\frac{\ell_p^2 \kappa}{2\pi} \delta S+\Omega \delta J+\Phi \delta Q+\mu \delta N,
\ee
where $S$ is given by (\ref{esta}) and $\mu=\kappa \ell_p^2 \sigma(\gamma)/(2\pi)$ is the chemical potential.
Semiclassical consistency follows from the fact that the above first law is exactly equivalent (as a simple calculation shows) to the
usual geometric first law $\delta M=\frac{\kappa}{2\pi} \delta (A/4)+\Omega \delta J+\Phi \delta Q$
for all values of the Immirzi parameter $\gamma$.
\section{Acknowledgements}
We would like to thank E. Wilson-Ewing for the careful reading of the short version of this comment. We also thank E. Frodden for discussions that concern the material of the appendix.

\begin{appendix}
\section{Reply to the rest}\label{more}

Most of the other points raised by A. Majhi
%, in consultation with P. Majumdar, 
are based on basic confusions. The difficulty with this second part is that it is much less organized than the first one: an enumeration of things in a chaotic way. 
So we shall only pinpoint the key conceptual problems that we could identify.

In the first paragraph of section III of \cite{idiot} A. Majhi claims that our energy is not the Komar energy. We agree with this; it is obviously so from our second paper \cite{ernesto} and explicitly stated in \cite{us}. It is neither the Komar-energy, nor the ADM-energy. It is the energy (the usual definition) measured by a family of preferred local stationary observers closely related §to ZAMOS. It is a quasi-local notion of energy that these preferred observers measure as the energy of the horizon. By no means it is the energy of the spacetime as a whole. We think we have adequately explained this in \cite{ernesto}. A precise explanation of the reason why this is a useful energy notion necessitated the careful argumentation presented in \cite{ernesto}. We encourage the reader to read our paper.

On the second paragraph in Section III of \cite{idiot}, the length scale $\ell$ is associated with the preferred family of observers we considered---it corresponds to the proper distance of these observers to the horizon. Alternatively, it is exactly equal to the inverse of the proper acceleration of our local observers. So its appearance in the energy measured by these observers is only natural. 
It's a coordinate independent quantity defining the observers; it is not also surprising that it is independent of the black hole parameters (for instance the observers decide to tune their spaceship engines at a certain fixed acceleration $1/\ell$, irrespective of the black hole spacetime). Only its expression in the Schwarzschild coordinates (e.g. the quantity $\epsilon$ in our paper) depends on the BH parameters. The only assumption related to the BH parameters is that $\ell<<\sqrt{A}$. 
An important thing is that $\ell$ is assumed to be the same before and after a perturbation of the BH to be described  by the local first law.
Once again this is a requirement in our definition of our local observers. One cannot object to this because this is simply an input in our construction (with well defined geometric and physical meaning). 
This input then leads to the novel structures we find.
Another crucial thing is that the parameter $\ell$ disappears from all the important results of \cite{us} (for instance it is not present in the summary of the main results presented at the end of Section \ref{uno} of the reply).

% On the third and fourth paragraphs of Section III of \cite{idiot}. The Energy for local observers at distance $\ell$ is 
% shown to be $E=A/(8\pi\ell)$ in \cite{}. The area is entirely a boundary operator whose spectrum is known exactly.
% We combine these two facts in our statistical mechanical treatment. The existence of the parameter $\ell$ in the energy
% is a label that tell us what are the observers to whom this energy means {\em energy}.  It can be seen as a regulator, a intermediate tool, etc.
% It is a feature of our model. The key fact is that this parameter disappears from the entropy and first law which is the central result of our work (see eqs. (\ref{esta}) and (\ref{pri})).

On Section IV: We consider the exact canonical partition function (not an expansion around the dominant configuration, as they incorrectly pointed out). Moreover, as described in Section {\ref{uno}}, our model boils down to a Boltzmann gas of non interacting particles with a definite single-particle energy levels. The statistical mechanics of such a system is absolutely standard and the comparison between various ensembles is also so. The thermodynamic stability issue is adequately dealt with in \cite{us} and the specific heat is positive. They did some incorrect calculation in an earlier comment (again errors are at a level of differential calculus) to arrive at a negative specific heat! 

% The only thing  that we neglect is the closure constraints. If one takes it into account this leads to logarithmic corrections to the entropy in the microcanonical 
% ensemble, in the canonical ensemble the effects are milder but it does not lead to any inconsistencies as A. Majhi seems to imply. The effect of this approximation is not important for our result.

On the statements in appendix A of \cite{idiot}. Here A. Majhi states that ``the energy flux associated to (our) family of observers $\sO$ is given by 
\be
\int_{\sW_{\sO}} \delta T_{ab} \chi^aN^b"
\ee
This is completely wrong and it is the source of all the rest of his confusions. This could be interpreted as the energy for a family of observers only if 
$\chi^a$ would be their four velocity, which is clearly not the case  as $\chi\cdot\chi\not=-1$. On the other hand, in contrast to what A. Majhi states in his comment, there is in no ambiguity in assigning 
\be
\int_{\sW_{\sO}} \delta T_{ab} u^aN^b
\ee
as the energy flux across the world-sheet $\sW_{\sO}$ as measured by $\sO$. More precisely, if $\sO$ would collect the in-falling matter by interposing a calorimeter the above quantity is exactly the amount of Jules measured by the calorimeter (this is all the energy falling into their system according to $\sO$ in the linearized setting where we are working). 
The fact that $\delta T_{ab}u^a$ is not conserved as a current does not prevent its physical interpretation presented above. 

Finally, it is not clear in what sense they say that the energy is not ``conserved". If they mean
that our energy depends on the geometry of the two surface (in our case completely fixed by the parameter $\ell$) then it is well-known 
in general relativity that quasi-local energies are not necessarily independent of the choice of the two surface (see e.g., J.D. Brown and J.W. York, Phys. Rev. D, 47 (1993) 1407).
It is very clear that our energy notion  does depend on $\ell$ (this is stated very explicitly in eq. (8) of \cite{us}, namely $E=A/(8\pi \ell)$); however, for each value of $\ell$ is has the unambiguous physical meaning explained above.  So this objection too is based on confusion.

\end{appendix}

\end{document}